\documentclass[twocolumn,superscriptaddress,nofootinbib,floatfix,aps,prb,citeautoscript,longbibliography]{revtex4-2}
\usepackage[utf8]{inputenc}
\usepackage[T1]{fontenc}
\usepackage{lineno}

\usepackage[colorlinks=true,linkcolor=blue,citecolor=blue,urlcolor=blue]{hyperref}
\usepackage{graphicx}%[draft]
\usepackage{xcolor}
\usepackage{xstring}
\usepackage{array}
\usepackage{dcolumn}
\usepackage{booktabs}
\usepackage{textgreek}
\usepackage{bm}
\usepackage{multirow}
\usepackage[version=4]{mhchem}
\usepackage{cleveref}
\usepackage{orcidlink}
\usepackage{float}
\newcommand{\etal}{{\emph et al.}}
\newcommand{\Tc}{$T_\text{c}$}

\newcommand{\bochum}{Research Center Future Energy Materials and Systems of the University Alliance Ruhr and Interdisciplinary Centre for Advanced Materials Simulation, Ruhr University Bochum, Universit\"atsstraße 150, D-44801 Bochum, Germany}

\newcommand{\jsnu}{Jiangsu Key Laboratory of Extreme Multi-Field Material Physics, School of Physics and Electronic Engineering, Jiangsu Normal University, Xuzhou 221116, China}

\newcommand{\cfm}{Centro de Fi\'sica de Materiales (CFM-MPC), CSIC-UPV/EHU, Manuel
de Lardizabal Pasealekua 5, 20018 Donostia/San Sebasti\'an, Spain}

\newcommand{\jilin}{Key Laboratory of Material Simulation Methods and Software of Ministry of Education and State Key Laboratory of Superhard Materials, College of Physics, Jilin University, Changchun 130012, China}

\begin{document}

\title {Predicted high-\Tc\ superconductivity above 130 K in metastable cubic \ce{MH4} compounds at ambient pressure}
\author{Xinxin Li} \thanks{These authors contributed equally}
%\email{Lixx@jsnu.edu.cn}
\author{Weishuo Xu} \thanks{These authors contributed equally}
%\email{xyz544256949@outlook.com}
\affiliation{\jsnu}
\author{Jingming Shi\orcidlink{0000-0003-3840-1988}}
%\email{jingmingshi@jsnu.edu.cn}
\affiliation{\jsnu}
\author{Hanyu Liu\orcidlink{0000-0003-2394-5421}}
%\email{hanyuliu@jlu.edu.cn}
\affiliation{\jilin} 
\author{Yue-Wen Fang\orcidlink{0000-0003-3674-7352}}\email{yuewen.fang@csic.es}
\affiliation{\cfm}
\author{Wenwen Cui\orcidlink{0000-0003-4097-1146}}
\email{wenwencui@jsnu.edu.cn}
\affiliation{\jsnu}
\author{Yinwei Li\orcidlink{0000-0002-9974-504X}}\email{yinwei\_li@jsnu.edu.cn}
\affiliation{\jsnu}
\author{Miguel A. L. Marques\orcidlink{0000-0003-0170-8222}}\email{miguel.marques@rub.de}
\affiliation{\bochum} 
\date{\today}

\begin{abstract}
Hydrides have long been considered promising candidates for achieving room-temperature superconductivity; however, the extremely high pressures typically required for high critical temperatures remain a major challenge in experiment. 
Here, we use high-throughput first-principles calculations to propose a class of high-\Tc~ambient-pressure superconductors with \ce{MH4} stoichiometry. These hydrogen-based compounds adopt the \textit{bcc} \ce{PtHg4} structure type, in which hydrogen atoms occupy the  one-quarter body-diagonal sites of metal lattices, with the metal atoms acting as chemical templates for hydrogen assembly. Through comprehensive first-principles calculations, we identify three promising superconductors, \ce{PtH4}, \ce{AuH4} and \ce{PdH4}, with superconducting critical temperatures of 84~K, 89~K, and 133~K, respectively, all surpassing the liquid-nitrogen temperature threshold of 77 K. The remarkable superconducting properties originate from strong electron-phonon coupling associated with hydrogen vibrations, which in turn arise from phonon softening in the mid-frequency range. Moreover, \textit{ab initio} molecular dynamics simulations demonstrate that \ce{PtH4} remains kinetic stable at 300 K over 20 ps, with no atomic diffusion observed. 
\end{abstract}  
\maketitle

\section{INTRODUCTION}

The quest for high-temperature superconductors has been a central focus in condensed matter physics. 
The metallic form of hydrogen, the lightest element, was historically proposed as a potential high-\Tc~superconductor by Ashcroft~\cite{ashcroft1968metallic}, building on the early theoretical work of  Bardeen, Cooper, and Schrieffer~\cite{bardeen1957theory,bardeen1957microscopic}.
However, metallic hydrogen is notoriously difficult to realize in experiment, even under pressures approaching 500~GPa~\cite{eremets2019semimetallic}. This challenge has redirected attention toward hydrogen-based compounds, where chemical precompression can significantly reduce the metallization pressure~\cite{ashcroft2004hydrogen}. A landmark example is H$_3$S, which was predicted to exhibit a \Tc~of 203~K at 200 GPa~\cite{li2014metallization,duan2014pressure} and was subsequently confirmed experimentally as a high-temperature superconductor~\cite{drozdov2015conventional}. Recent advances in high-pressure synthesis have further expanded the family of hydride superconductors to include compounds such as LaH$_{10}$~\cite{drozdov2019superconductivity,somayazulu2019evidence}, YH$_6$ and YH$_9$~\cite{liu2017potential,peng2017hydrogen}, CaH$_6$~\cite{wang2012superconductive,ma2022high,li2022superconductivity}, as well as ternary hydrides including (La,Y)H$_6$~\cite{semenok2021superconductivity}, (La,Ce)H$_9$~\cite{chen2023enhancement,bi2022giant}, (La,Al)H$_{10}$~\cite{chen2024high}. In particular, a recent experimental study of LaSc$_2$H$_{24}$ reported a \Tc~of 298 K at 260 GPa~\cite{song2025exp-298K-La-Sc-H}, providing evidence of room-temperature superconductivity in reality.

Extreme pressures pose, however, significant experimental challenges, driving efforts to identify stable superconductors at lower or even ambient pressure.
Representative examples include perovskites~\cite{DANGIC2025100043,cerqueira2024searching}, inverted perovskites~\cite{cerqueira2024searching}, and perovskite-derived SM$_2$-TM-H$_6$~\cite{sanna2024prediction,cerqueira2024searching,gao2025maximum,dolui2024feasible,sanna2025gnome-hydrides,MgAlFe6-MIT-JuLI2025}, and other families with high-symmetry structures~\cite{Miguel2026March-CMT-superC}. In perovskite systems, compounds such as \ce{KInH3}, \ce{SrAuH3}, and \ce{SrZnH3} have been predicted to exhibit high \Tc\ values up to 132~K at ambient pressure~\cite{li2024theoretical,cerqueira2024searching}. Du et al. predicted a series of high-\Tc\ superconductors below 10~GPa, with their best candidate having a maximum \Tc\ of 170~K at 4 GPa, although these phases are thermodynamically metastable~\cite{du2024high}. 
In spite of being a binary hydride, the cubic crystal structure of the \ce{M4H} template adopts a perovskite-like framework, characterized by the same space group $Pm\bar{3}m$ and lattice topology as the ideal cubic perovskite.
Among such binary cubic \ce{M4H} hydrides, \ce{Al4H} and~\ce{Ag4H} have been predicted to show \Tc~of 54 and 63~K, respectively~\cite{he2023phonon,tian2025ambient,cerqueira2024searching}.
Compared to the aforementioned perovskites, the inverted perovskites tend to display moderate \Tc. The machine-learning-aided searches have uncovered nearly 70 inverted perovskites with \Tc\ ranging from 5 to 22 K~\cite{hoffmann2022superconductivity,cerqueira2024searching}, while the current record is held by \ce{MgHCu3} which shows superconductivity at 42~K~\cite{tian2024few}.
For the SM$_2$-TM-H$_6$ family with vacancy-ordered perovskite structure~\cite{sanna2025gnome-hydrides}, the critical temperature of superconducting Mg$_2$IrH$_6$ is predicted to reach more than 77~K at ambient pressure~\cite{sanna2024prediction,dolui2024feasible}, and that of electron-doped Mg$_2$PtH$_6$ can even surpass 100~K~\cite{sanna2024prediction}. More recently, machine-learning assisted prediction has identified more high-\Tc\ SM$_2$-TM-H$_6$ hydrides at atmospheric pressure such as \ce{Li2AuH6} (\Tc$\sim$116 K) and \ce{Li2AgH6} (\Tc$\sim$108~K)~\cite{ouyang2025high,gao2025maximum}. Apart from these perovskite and perovskite-derived structures, the cubic \ce{X4H15} systems offer another route to achieving ambient-pressure superconductivity, where hole doping can be strategically designed to both boost the \Tc\ values up to 50~K and improve the thermodynamic stability~\cite{gao2025enhanced}.

Recently, the ''chemical template concept'' was proposed to successfully elucidate the mechanism of high-\Tc\
superconductivity in hydrides  such as \ce{CaH6}, \ce{YH6},  \ce{CeH9}, and \ce{LaH10}~\cite{sun2023chemical}. In these systems, interstitial quasi-atoms (ISQs) exhibit an excellent structural match with the extended superhydrogen network embedded in the parent body- or face-centered cubic (\textit{fcc}) metal lattice~\cite{sun2023chemical}. 
Building on this framework, some superconductors have been predicted recently. For example, incorporation of \ce{H2} units into \textit{fcc} NaK framework led to the formation of \ce{NaKH12}, which displays a critical temperature of 246~K at 60 GPa~\cite{zhao2025unlocking}. In another case, hydrogen intercalation in the interstitial sites of sodium ion layers, combined with electronic matching to a honeycomb boron lattice, yielded a superconductor with a maximum \Tc\ of 63~K at ambient pressure~\cite{zhao2024high}. More recently, by integrating the “chemical template effect” with machine learning algorithms, Miao \etal~identified 13 new structural prototypes and 31 thermodynamically stable metal superhydrides at pressures ranging from 100--300 GPa, among which 19 exhibiting \Tc\ above 100 K~\cite{sun2025batch}. 
These examples demonstrate that the chemical template concept provides a useful approach for designing new superconductors.

Here, the $bcc$ structure is chosen as the starting point of our electron localization function (ELF)-guided design strategy for both physical and practical reasons. First, $bcc$ lattices are energetically favorable for many elemental metals under ambient or moderately compressed conditions and are widely used as reference structures in high-pressure and hydrogen-rich materials searches. Second, the $bcc$ lattice exhibits well-defined and spatially separated maxima in the ELF within its interstitial regions (see Fig. S1 in Supplementary Materials), making it particularly suitable for identifying distinct ELF "islands" that can act as favorable sites for hydrogen incorporation.

Furthermore, within the framework of the chemical template concept, the high-\Tc\ \ce{CaH6} and \ce{YH6} can be characterized as a combination of a \textit{bcc} metal (e.g. Ca) with a hydrogen network where the hydrogen atoms occupy tetrahedral interstitial sites $12d$ $(\tfrac{1}{2}, 0, \tfrac{1}{4})$ of the \textit{bcc} lattice~\cite{sun2023chemical}. Such \textit{bcc} metals are hereafter referred to as type I. In contrast, our analysis of the electron localization function (ELF) for a range of simple \textit{bcc} metals reveals that, in some other cases such as Pt (denoted as type II), the interstitial electrons can occupy positions at around the one-quarter points along the body diagonals $8c$ $(\tfrac{1}{4}, \tfrac{1}{4}, \tfrac{1}{4})$. The comparison of the ELF between the type-I and type-II metals are shown in Supplementary Fig. S1.  Incorporation of hydrogen atoms into these one-quarter positions gives rise to a unexplored PtHg$_4$-type structure, \ce{MH4}.

In this study, guided by the concept of chemical template, high-throughput first-principles calculations have been used to predict a novel family of metal hydrides, \ce{MH4}, with a PtHg$_4$-type structure ~\cite{haussermann2001bonding} where hydrogen atoms are bonded in the framework of \textit{bcc} metals. In particular, we identify three high-\Tc\ superconducting hydrides at ambient pressure, \ce{PtH4}, \ce{AuH4}, and \ce{PdH4}, with superconducting critical temperatures of 84~K, 89~K, and 133~K, respectively, all surpassing the boiling point of liquid nitrogen (77 K). In these \ce{MH4} superconductors, dynamical stability and enhanced superconductivity are closely related to the metallic lattice. 
Our work opens a theoretical avenue to the exploration of plausible structures with potentially novel properties, and advances the search for high-\Tc\ superconductivity.

\raggedbottom
\begin{figure*}[htp]
\centering
  \includegraphics[width=1\linewidth,angle=0]{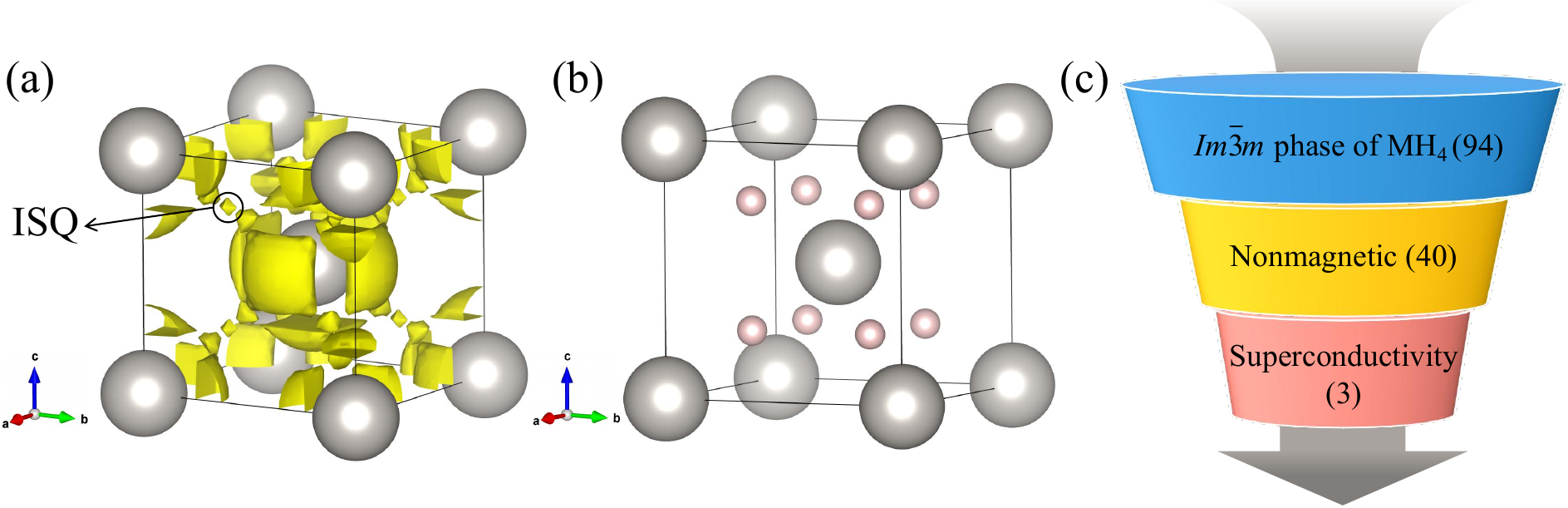}
  
  \caption{\label{fig:1} (a) Three dimensional electron localization function of the \textit{bcc} lattice formed by the M atoms (isosurface = 0.326). The circle highlights the interstitial quasi-atoms (ISQ). (b) The structure of MH$_4$ at ambient pressure. The M and H atoms are colored in gray and pink, respectively. (c) Schematic of our screening workflow. Panels (a) and (b) were produced with VESTA~\cite{Momma:ko5060}.}
\end{figure*}

\section{Computational Details}

Calculations were performed of the entire family of PtHg$_4$-type \ce{MH4} compounds,
where M represents all the metal atoms in periodic table.  Meanwhile,\ce{PdH4} and \ce{PtH4} are predicted using the CALYPSO code~\cite{PhysRevB.82.094116,WANG20122063,GAO2019301,10.1063/5.0074677} based on a particle swarm optimization algorithm at different pressure.
Geometry relaxations, total energy  calculations, and electronic properties were performed using the  Vienna $Ab$ $initio$ Simulation Package (VASP)~\cite{kresse1996efficient} using the projector augmented-wave (PAW) method. 
The exchange-correlation functional of density-functional theory was approximated by the generalized gradient approximation of Perdew, Burke, and Ernzerhof (PBE-GGA)~\cite{perdew1997generalized}. 
The ion-electron interaction was described by PAW potentials with $5d^96s^1$, $4d^95s^1$, $5d^{10}6s^1$ and $1s^1$ electrons configurations as valence states for Pt, Pd, Au and H atoms, respectively. The cutoff energy for the expansion of the wave functions in the plane wave basis was set to 600~eV, with a grid of spacing $2\pi\times0.03$~\AA$^{-1}$ for the electronic Brillouin zone integration~\cite{tang2009grid}. In the high-throughput search, the dynamical stability was assessed by using the finite displacement and supercell method implemented in {\sc phonopy} code~\cite{togo2015first}.
$Ab$ $initio$ molecular dynamics simulations were performed in canonical (NVT) ensemble using a Nos\'e-Hoover thermostat~\cite{evans1985nose}
at 300 K. A time step of 1~fs and a total simulation time of 20~ps were used. The simulations employed a $3\times3\times3$ supercell of the conventional unit cell containing 270 atoms.

To evaluate the superconducting properties, electron-phonon coupling (EPC) calculations were performed within Eliashberg theory for phonon-mediated superconductors under the Migdal approximation~\cite{li2025extended,han2025pressure}. Phonon spectra and EPC constants were obtained by density functional perturbation theory (DFPT) method using {\sc quantum espresso}~\cite{giannozzi2009quantum,giannozzi2017advanced,giannozzi2020quantum}. PAW pseudopotentials from the PSlibrary~\cite{dal2014pseudopotentials} were used together with a kinetic energy cutoff of 90~Ry.  In order to further assess the reliability of the PAW pseudopotentials employed in {\sc vasp} and {\sc quantum espresso}, we compared the Birch--Murnaghan equations of state obtained using the PAW potentials with those calculated using the full-potential linearized augmented plane-wave (LAPW) method with local orbitals, as implemented in {\sc wien2k}~\cite{BLAHA1990399}. As shown in Fig.S2, the excellent agreement between the two approaches provides an independent validation of the reliability of the adopted PAW potentials.

The Brillouin zone was sampled using a $24\times24\times24$ $k$-point mesh and a $6\times6\times6$ $q$-point mesh for EPC calculations. A Gaussian smearing of 0.04~Ry was employed in evaluating EPC properties. The electron-phonon coupling properties were calculated using Gaussian broadenings ranging from 0 to 0.06~Ry with a step of 0.002~Ry and a $k$-point grid of $N \times N \times N$ ($N = 18, 24, 30,$ and $36$). Our test calculations find that a $24 \times 24 \times 24$ $k$-point mesh with a Gaussian broadening of 0.04 Ry yields well-converged values of $\lambda(q)$ (see Fig. S4). We believe that our choices for the $k$-mesh as well as the choice for the Gaussian broadening, provide an appropriate balance between accuracy and computational efficiency, making them suitable for calculations across the studied compounds in our study. Therefore, these parameters were used to calculate the reported values of \Tc.
The superconducting critical temperature was calculated by using Allen-Dynes equation~\cite{allen1975transition} with a standard Coulomb pseudopotential $\mu^{*}=0.10$~\cite{gao2025maximum,Gao-PRB2024-AcBeH,PRB2025-BeBaH8}:
\begin{equation}
  T_\text{c} = f_{\text{1}}f_{\text{2}}\frac{\omega_{\text{log}}}{1.2} \text{exp}\left[\frac{-1.04(1+\lambda)}{\lambda-\mu^*(1+0.62\lambda)}\right] ,
  \end{equation}
where $\omega_{\text{log}}$ is the logarithmic average frequency, and $\lambda$ is the EPC parameter.
The factors, 
$f_{\text{1}}$ and $f_{\text{2}}$ are strong-coupling correction and shape correction factors, respectively.

The isotropic Migdal-Eliashberg equations take the form~\cite{allen1983theory}:

%%%%%%%%%%%%%%%%%%%%%%%%%
\begin{equation}
\begin{aligned}
[1 - Z(i\omega_n)] i\omega_n
= & -\frac{\pi}{\beta} \sum_{n'}
\frac{Z(i\omega_{n'}) i\omega_{n'}}{\Xi(i\omega_{n'})} \\
& \times \int
\frac{2\omega a^2 F(\omega)}
{(\omega_n - \omega_{n'})^2 + \omega^2}
\, d\omega
\end{aligned}
\end{equation}

\begin{equation}
\begin{aligned}
\Phi(i\omega_n)
= & \frac{\pi}{\beta} \sum_{n'}
\frac{\Phi(i\omega_{n'})}{\Xi(i\omega_{n'})} \\
& \times \left[
\int
\frac{2\omega a^2 F(\omega)}
{(\omega_n - \omega_{n'})^2 + \omega^2}
\, d\omega
- N_F \bar{V}_C
\right]
\end{aligned}
\end{equation}

\begin{equation}
\Phi(i\omega_n) = \Delta(i\omega_n) Z(i\omega_n)
\end{equation}

\begin{equation}
\Xi(i\omega_n)
= \sqrt{[Z(i\omega_n)\omega_n]^2 + [\Phi(i\omega_n)]^2}
\end{equation}
where $\Delta(i\omega_n$) and $Z(i\omega_n$) represent superconducting gap and the renormalization function, $N_F$ is the density of electronic states at the Fermi level, and $\omega_n$ = $(2n+1)\pi/\beta$ (with $n$ integer) are the fermion Matsubara frequencies.

It should be mentioned that conventional Migdal-Eliashberg theory relies on the Migdal's theorem~\cite{Migdal1958InteractionJETP} that assumes the phonon energy scale is much smaller than the electronic energy scale.
In practice, the vertex corrections are neglected in Migdal-Eliashberg theory when the expansion parameter $\lambda (\hbar \omega_0 / E_{\rm F})$ is much smaller than the unity,
where ${\omega_0}$ and $E_{\rm F}$ are characteristic phonon frequency and Fermi energy, respectively. 
In some superconductors, such as high-\Tc~\ce{H3S}, this parameter is not small. However, a recent systematic study on \ce{H3S} at 200 GPa has shown that these corrections remain controlled and the variation of \Tc~is mild upon the inclusion of vertex corrections (from 189 to 178 K), with excellent agreement with the experimental \Tc~of around 175-185 K at 200$\pm$5 GPa~\cite{Mishra2025-npj-comput-H3S-vertex}.
In addition, the constant DOS approximation is employed in conventional  Migdal-Eliashberg theory and it may lead to quantitative deviations in the predicted \Tc~in systems with strongly energy-dependent DOS, such as \ce{H3S} and \ce{Mg2IrH6} where there is a peak of DOS near the Fermi level. These effects on \Tc~are, however, generally not dramatic, and can be properly accounted for in the density functional theory for superconductors (SCDFT) and full-bandwidth Eliashberg's theory, in which the energy dependence of DOS is explicitly included~\cite{Mishra2025-npj-comput-H3S-vertex,Arita-H3S-2015PRB-SCDFT,sanna2024prediction}. 
For example, \ce{Mg2IrH6} shows a \Tc~of 77 K in conventional Migdal-Eliashberg equation and is slightly reduced to 71 K in SCDFT~\cite{sanna2024prediction}.
In this regard, our predictions based on the conventional Migdal-Eliashberg equation are expected to remain quantitatively reliable.

\section{RESULTS AND DISCUSSION}

Unlike the valence electrons in type I metals (e.g., Ca, Y, and La), which tend to localize near the tetrahedral interstitial sites in the crystal lattice~\cite{sun2023chemical}, the electrons in type II metals (e.g., Pt and Rh) exhibit pronounced delocalization around the one-quarter body-diagonal position of the \textit{bcc} lattice, $8c$ $(\tfrac{1}{4}, \tfrac{1}{4}, \tfrac{1}{4})$. This feature is clearly demonstrated by the ELF shown in Fig.~\ref{fig:1}(a) and Supplementary Figs. S1 and S3.
 In contrast to the hydrogen networks observed in superconducting hydrides such as \ce{CaH6} and \ce{YH6}, our calculations reveal that the intentional incorporation of hydrogen atoms into the electron-rich interstitial regions of these metals leads to the stabilization of a novel \ce{MH4} structural framework. The resulting \ce{MH4} phases crystallize in the cubic structure with space group $Im\bar{3}m$, adopting the PtHg$_4$-type prototype. Structure-equivalent phases such as \ce{CrGa4} and \ce{MnGa4} have been experimentally synthesized and remain stable up to $\sim$100 kbar~\cite{haussermann2001bonding}. In \ce{MH4}, the M atoms occupy the Wyckoff $2a$ (0,0,0) positions corresponding to the corners and body center of the cubic cell, while the H atoms reside at the Wyckoff $8c$ $(\tfrac{1}{4}, \tfrac{1}{4}, \tfrac{1}{4})$ sites. Each M atom is thus coordinated by eight equivalent H atoms in a \textit{bcc}-like geometry, as illustrated in Fig.~\ref{fig:1}(b).

\begin{figure*}[htp]
\centering
  \includegraphics[width=0.8\linewidth,angle=0]{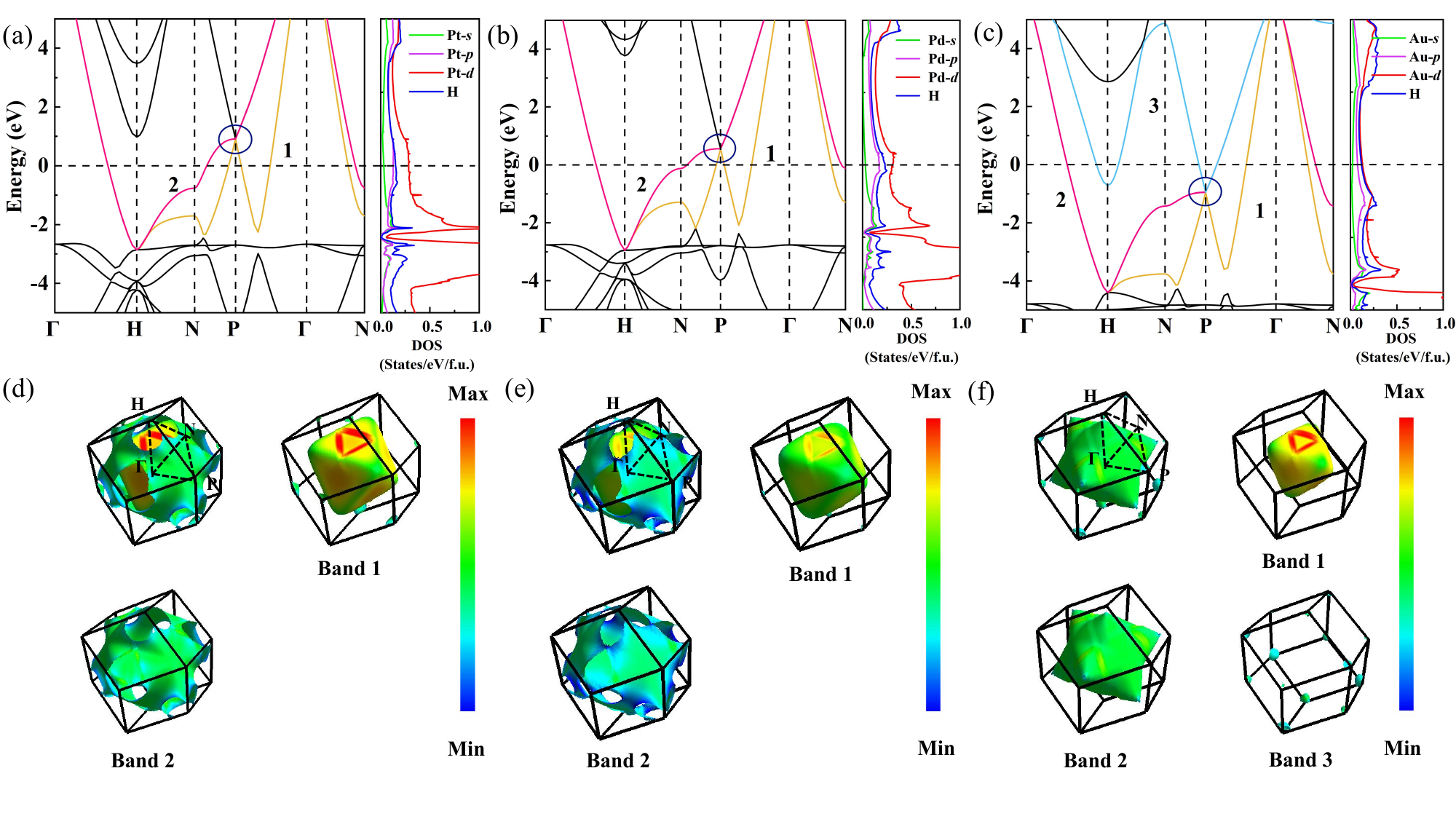}
  \caption{\label{fig:2} Calculated band structures and projected density of states (PDOS) of (a-c) PtH$_4$, \ce{PdH4} and AuH$_4$ at 0 GPa. The circles highlight the triply degenerate (spinless) point where three energy bands meet together. The colored lines, labeled ``1'', ``2'' or ``3'', in the band structure are used solely to guide the eye toward the bands crossing the Fermi level. In contrast, in the PDOS diagram, the colors correspond to different orbital or elemental characters.  (d-f) Fermi surfaces of PtH$_4$, \ce{PdH4} and AuH$_4$, colored with respect to the Fermi velocity $\left\langle{v}\right\rangle$ (10$^5$ m/s). The color bars use the same scale of $\left\langle{v}\right\rangle$.
  }
\end{figure*}

High-throughput first-principles screening was carried out for all \ce{MH4} compounds with M spanning all the  metal elements. The computational workflow of materials screening is schematically summarized in Fig.~\ref{fig:1}(c).
This procedure identified three dynamically stable compounds:  \ce{PdH4}, \ce{PtH4}, and \ce{AuH4} (Fig. S5). In addition, these exhibit superconductivity with a \Tc~exceeding the boiling point of liquid nitrogen. To clarify the underlying superconducting mechanism, we first calculated their electronic band structures and densities of states (DOS), as shown in Figs.~\ref{fig:2}(a-c). Due to the chemical similarity of Pt and Pd, both compounds exhibit analogous electronic features. The $d$ orbitals of the metal atoms dominate the DOS at the Fermi level (${E_{\rm F}}$), contributing 0.18, 0.38 and 0.15 eV$^{-1}$ per primitive cell and accounting for 45\%, 32\% and 40\% of the total DOS in \ce{PtH4}, \ce{PdH4} and \ce{AuH4}, respectively. There is a significant overlap in partial DOS between metal and H atoms in the studied energy window from -4.5 to 4.5 eV, indicating that each pair of M--H is strongly hybridized.

We further investigate the spin-orbit coupling (SOC) effect on the band structures of \ce{MH4} compounds (see Fig. S6). The three bands crossing the Fermi level remain almost unchanged upon SOC inclusion. Consequently, SOC does not significantly modify the Fermi surface topology, Fermi velocities, or density of states at the Fermi level.

As seen from the electron bands in Figs.~\ref{fig:2}(a,b) and the Fermi surfaces (FS) in Figs.~\ref{fig:2}(d,e), two bands, labeled \#1 and \#2,  cross ${E_{\rm F}}$ in both \ce{PtH4} and \ce{PdH4}, forming two FS. Bands \#1 and \#2 are degenerate along the high-symmetry line from $\Gamma$ to H in the first Brillouin zone. Band \#1 exhibits a regular octahedral FS centered at the $\Gamma$ point, whereas Band \#2 consists of six interconnected, planar-quadrilateral-like Fermi sheets with a clover-shaped opening at the H point. 

\raggedbottom
\begin{figure*}[htp]
\centering
  \includegraphics[width=0.85\linewidth,angle=0]{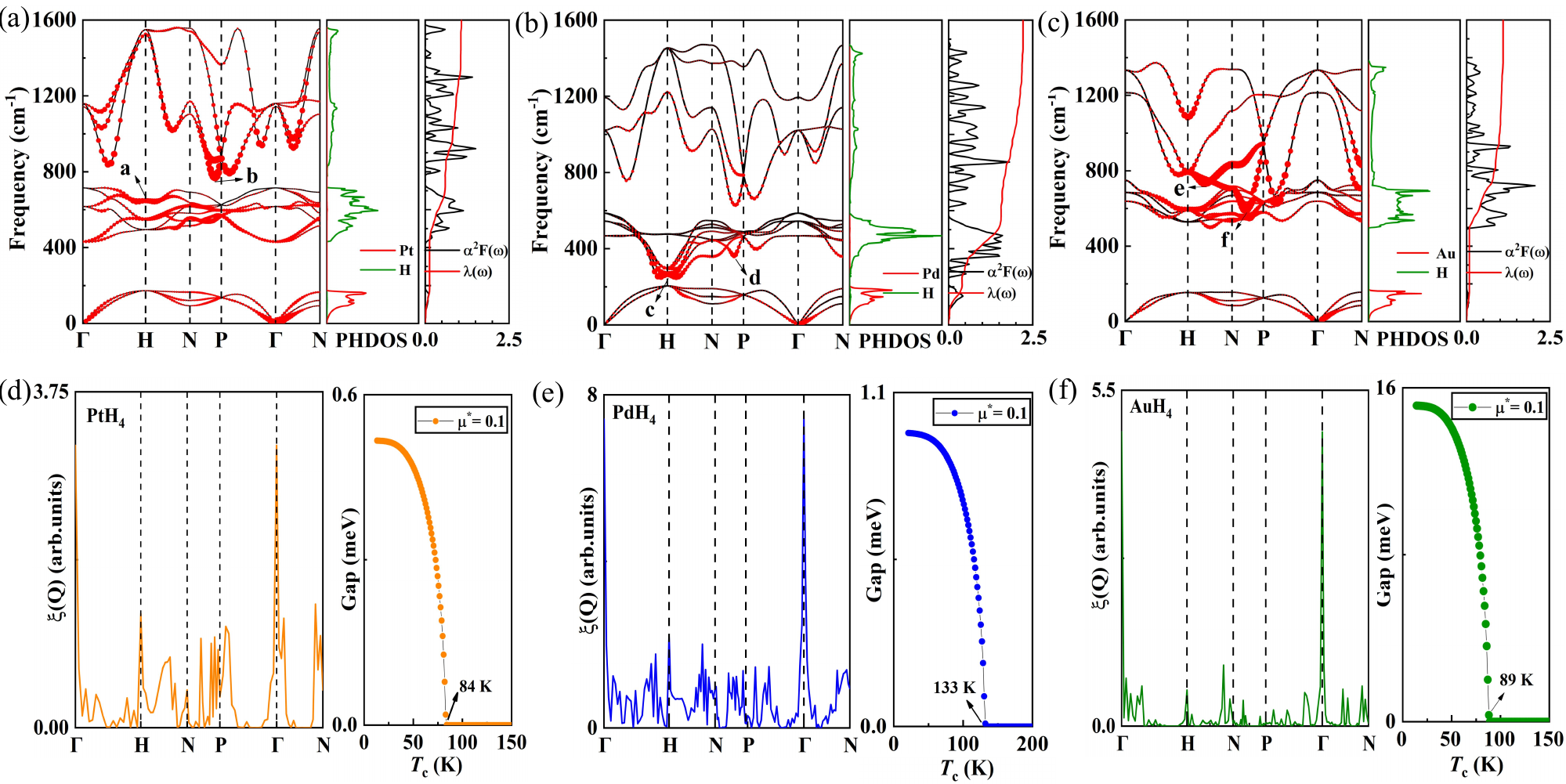}
  \caption{\label{fig:3} Calculated phonon band structures, phonon density of states (PHDOS), electron-phonon coupling coefficient $\lambda(\omega)$ and Eliashberg spectral function $\alpha^2F(\omega)$ for (a) PtH$_4$, (b) PdH$_4$, and (c) \ce{AuH4} at 0 GPa. The red solid circles show the phonon linewidth with a radius proportional to the EPC strength.  The arrows denote the significant  vibration modes in phonon dispersion curves. We also present the calculated nesting function $\xi$(Q) at 0 GPa along some particular $q$ trajectories and isotropic superconducting gap obtained by numerically solving the isotropic Migdal-Eliashberg equations for (d) \ce{PtH4}, (e) \ce{PdH4}, (f) \ce{AuH4}, respectively.
  }
\end{figure*}

Since Au possesses one additional valence electron compared to Pt and Pd~\cite{rahm2017ternary}, Fig.~\ref{fig:2}(c) shows that the Fermi level in the band structure exhibits a distinct upward shift. Consequently, in the case of \ce{AuH4}, three bands go across ${E_{\rm F}}$, labeled as \#1, \#2, and \#3, giving rise to the FS illustrated in Figs.~\ref{fig:2}(f). Band \#1, similar to that in either \ce{PtH4} or \ce{PdH4}, forms a regular octahedral FS centered at the $\Gamma$ point, whereas Band \#2 generates a cuboidal FS with smoothly rounded faces and sharply protruding vertices. In contrast, Band \#3 produces small Fermi pockets located at the H and P points of the Brillouin zone. Notably, the average Fermi velocity of \ce{PdH4} is lower than those of both \ce{PtH4} and \ce{AuH4}, potentially suggesting stronger EPC and enhanced superconductivity~\cite{kiss2007charge}, considering $N(E_F)$ inversely related to the average Fermi velocity.

To clarify the nature of the M–H interaction, we have performed additional crystal orbital Hamilton population (COHP) analysis and evaluated the corresponding integrated COHP (ICOHP) values, which explicitly quantify the bonding-antibonding contributions arising from orbital overlap between metal and hydrogen atoms.
As shown in Fig. S8, the ICOHP values for Pt-H and Au-H bonds are -1.69 eV/pair and -1.37 eV/pair, respectively, both of which are significantly more negative than that of the Pd-H bond (-0.75 eV/pair). These results clearly indicate stronger orbital hybridization and bonding interactions for Pt-H and Au-H compared to Pd–H, thereby supporting our original conclusion regarding the relative bond strengths.

To further study the lattice dynamics and electron-phonon coupling properties, we used the DFPT method to calculate the phonon dispersion curves, phonon density of states (PHDOS), and Eliashberg spectral function $\alpha^2F(\omega)$ for \ce{PtH4}, \ce{PdH4} and \ce{AuH4}, as shown in Figs.~\ref{fig:3}(a-c), respectively.
The values of the EPC constant ($\lambda$), logarithmic average phonon frequency ($\omega_\text{log}$), and \Tc\ are included in Table~\ref{tab:tc_values}. % reveal significant differences in the \Tc~among the studied compounds. 
Using the Allen-Dynes equation with a Coulomb pseudopotential of 0.10, the ambient-pressure \Tc\ for superconductivity are 75~K for \ce{PtH4},  146~K for \ce{PdH4}, and 79~K for \ce{AuH4}. 
We also evaluated the~values of \Tc\ by solving self-consistently the isotropic Migdal-Eliashberg equations~\cite{eliashberg1960interactions}, which yields \Tc\ values of 84~K, 133~K, and 89~K for \ce{PtH4}, \ce{PdH4}, and \ce{AuH4}, respectively.

\begin{table*}
\centering
\caption{The calculated EPC parameter ($\lambda$), logarithmic 
average phonon frequency ($\omega_{\log}$), and estimated $T_c$ 
for MH$_4$ (M = Pd, Pt, and Au) at ambient pressure using the 
Allen-Dynes (AD) equations~\cite{allen1975transition} or by 
solving the Migdal-Eliashberg (ME) equations with 
$\mu^* = 0.1$, 0.13, and 0.15~\cite{eliashberg1960interactions}.}
\label{tab:tc_values}
\renewcommand{\arraystretch}{1.3}

\begin{ruledtabular}
\begin{tabular}{cccccc}
Hydride & $\lambda$ & $\omega_{\log}$ (K) & $\mu^*$ 
        & \multicolumn{2}{c}{$T_c$ (K)} \\
\cline{5-6}
        &      &     &      & AD  & ME  \\
\hline
PtH$_4$ & 1.08 & 892 & 0.10 & 75  & 84  \\
        &      &     & 0.13 & 55  & 79  \\
        &      &     & 0.15 & 43  & 62  \\
\hline
PdH$_4$ & 2.21 & 609 & 0.10 & 146 & 133 \\
        &      &     & 0.13 & 127 & 129 \\
        &      &     & 0.15 & 115 & 83  \\
\hline
AuH$_4$ & 1.14 & 873 & 0.10 & 79  & 89  \\
        &      &     & 0.13 & 59  & 83  \\
        &      &     & 0.15 & 47  & 77  \\
\end{tabular}
\end{ruledtabular}
\end{table*}

As seen in Fig.~\ref{fig:3}, the phonon dispersions of the three \ce{MH4} compounds can be roughly grouped into low-, medium-, and high-frequency regions based on the frequency distribution, where the low-lying phonons are dominated by metal atoms, and the latter two regions are exclusively from H atoms. 
In the cases of \ce{PtH4} and \ce{PdH4}, these three frequency regions are well separated.
For \ce{PtH4}, the three vibrational regions contribute 12\%, 44\%, and 44\% to the total electron-phonon coupling constant ($\lambda$ = 1.08). In contrast, for \ce{PdH4}, the gap between the low- and medium-frequency branches narrows significantly due to a prominent phonon softening at the H point. The softened modes at 250 $\sim$ 580 cm$^{-1}$  contributes  45\% to the EPC strength, leading to a much larger $\lambda$ value of 2.21.

Unlike the frequency distributions of \ce{PtH4} and \ce{PdH4}, the intermediate-frequency phonon modes touch the high-frequency modes in \ce{AuH4}. %and contribute 74\% to the total $\lambda$ 1.14  
The intermediate-frequency phonon modes (490 $\sim$ 720 cm$^{-1}$) contribute 55\% ($\lambda$ = 0.63) to the total $\lambda$ 1.14, and the high-frequency modes (720 $\sim$ 1380 cm$^{-1}$) are responsible for 37\% ($\lambda$ = 0.42) of the total $\lambda$  1.14 in \ce{AuH4}.
 %\edit{The metallic Au atoms contribute 20\% to the total PHDOS} \yw{(are you sure here it is PHDOS, not the $\lambda$?)}, \edit{while hydrogen accounts for the remaining 80\%} (\cww{Here, it is not right, not the pHDOS nor the $\lambda$. The student understood incorrect,  and we just delete it. }
The computational results reveal that the H vibrations dominate the Eliashberg spectral function $\alpha^2F(\omega)$, providing the primary contribution to the total EPC strength.  
For the phonon modes showing significant phonon linewidths in Figs.~\ref{fig:3}(a-c), their detailed vibration patterns  are shown in Fig. S9. All these vibration modes are dominated by the hydrogen atoms, indicating their crucial role in phonon softening and electron-phonon coupling.

Furthermore, the potential for Fermi surface nesting along the high-symmetry path was evaluated using the nesting function~\cite{tse2007novel}: 
\begin{equation}
    \xi(\bm{Q}) = \frac{1}{N} \sum_{\bm{k}, i, j} \delta(\varepsilon_{\bm{k}, i} - \varepsilon_{\bm{F}})\, \delta(\varepsilon_{\bm{k} + \bm{Q}, j} - \varepsilon_{\bm{F}}),
\end{equation}
where $\varepsilon_{\bm{k}, i}$ is the Kohn-Sham eigenvalue, and $i$, $j$ are the indices of energy bands, $N$ is the number of the $k$ points, and $\varepsilon_{\bm{F}}$ is the Fermi Energy.  As shown in the top panel of Fig.~\ref{fig:3}(d), the nesting function of \ce{PtH4} exhibits pronounced peaks along the H–N and N–P directions, corresponding to the phonon softening observed in Fig.~\ref{fig:3}(d). For \ce{PdH4}, $\xi(\bm{Q})$ exhibits several distinct peaks along the $\Gamma$–H and H–N paths [middle panel in Fig.~\ref{fig:3}(e)], which can be attributed to the Kohn anomalies near the H point [Fig.~\ref{fig:3}(b)]. These anomalies give rise to pronounced phonon softening~\cite{kasinathan2006superconductivity}. In \ce{AuH4}, clear peaks are also present along the H-N direction. The large phonon linewidths near H-N further account for the strong peak observed along this path.
These results of FS nesting, in combination with the analyses of phonon properties,  are responsible for the strong EPC and high \Tc~in  MH$_4$, in which the high \Tc~values are further confirmed by solving the isotropic Migdal-Eliashberg equations [Fig.~\ref{fig:3}(d)(e)(f)]. 

\begin{figure*}[htp]
\centering
  \includegraphics[width=0.85\linewidth,angle=0]{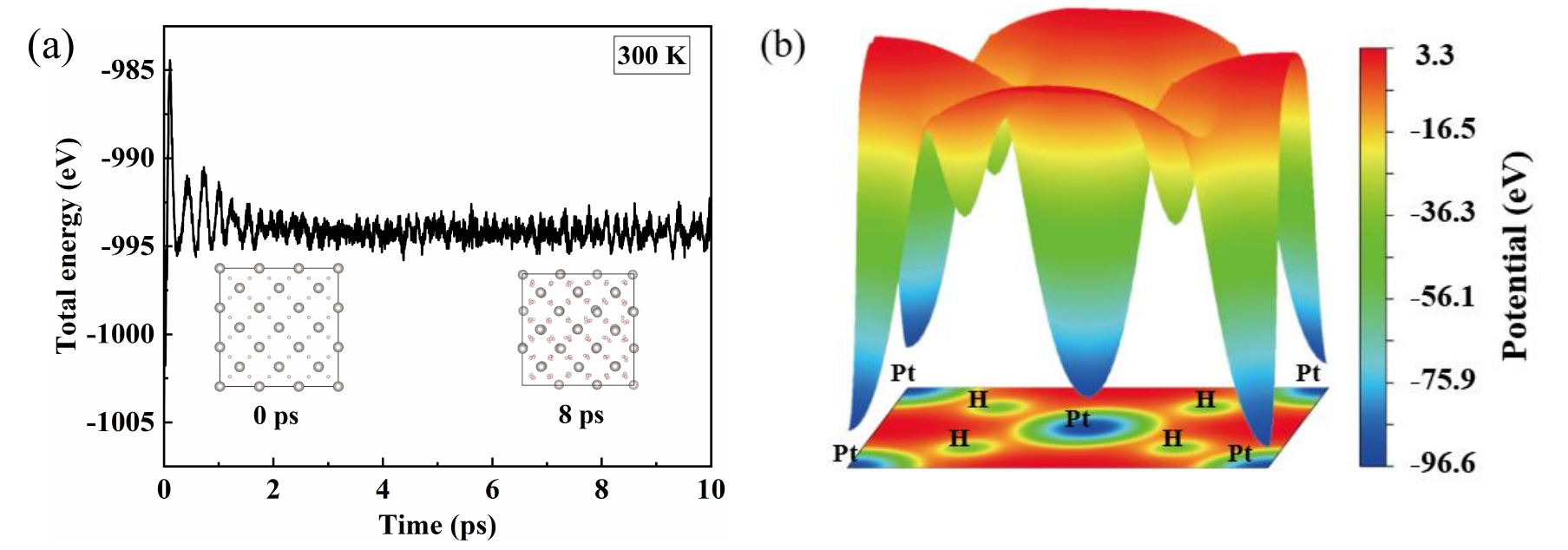}
  \caption{\label{fig:md} (a) Ab initio molecular dynamics simulation of PtH$_4$ at 0 GPa and 300 K. The inserts are snapshots of structures at 0 ps and 8 ps simulations, respectively. (b) Calculated electrostatic potential on the Pt-H plane. 
%   \yw{If I am not wrong, the y-label in (a) should be "Total energy" rather than "Total free energy"}
  }
\end{figure*}

 We further evaluate the dynamical stability of \ce{MH4} by performing ab initio molecular dynamics simulations at 300 K, and the corresponding results are presented in Fig.~\ref{fig:md}(a) and S10. In contrat to \ce{PtH4}, both \ce{PdH4} and \ce{AuH4} exhibit pronounced hydrogen diffusion and/or structural decomposition within a relatively short simulation time, indicating limited short-time kinetic stability at 300 K. Similar behavior has also been reported for \ce{Li2AuH6} and \ce{Li2AgH6}~\cite{ding2026kinetic}, which provide representative examples of hydrides with high \Tc\ values at ambient pressure  but exhibit kinetic instability on MD simulations.
 
In contrast, the supercell structure of \ce{PtH4} is well preserved without any structural collapse. As shown in Fig.~\ref{fig:md}(a), atomic displacements remain confined to small thermal fluctuations around equilibrium positions during a 20~ps trajectory, and the total energy exhibits only minor variations after equilibration. These results indicate robust short-time finite-temperature stability of \ce{PtH4} at 300 K. Therefore, although \ce{PdH4} has the highest predicted \Tc, \ce{PtH4} is a more appropriate representative for assessing the finite-temperature stability of the proposed \ce{MH4} structure. 

We note that a 10--20 ps AIMD trajectory primarily probes short-time local stability and cannot definitively exclude long-time-scale hydrogen diffusion, decomposition, or structural transformations.  This limitation is particularly important for hydrogen-rich systems, as recently demonstrated for \ce{Li2AuH6} and \ce{Li2AgH6}~\cite{ding2026kinetic}, where PIMD simulations revealed kinetic instability despite apparently stable AIMD trajectories over comparable time scales. Although long-timescale hydrogen diffusion or dehydrogenation processes are beyond the accessible simulation window, the absence of spontaneous lattice collapse and short-range hydrogen rearrangement on the picosecond timescale demonstrates the local dynamical stability at 300 K. 
 
 Furthermore, to provide an additional assessment of the finite-temperature dynamical stability of \ce{PtH4}, we have employed the temperature-dependent effective potential (TDEP) method to calculate its effective phonon spectrum at 300 K. The absence of imaginary frequencies further supports the dynamical stability of \ce{PtH4} at finite temperature. Thus, the AIMD results, together with the phonon and TDEP calculations, consistently support the short-time finite-temperature stability of \ce{PtH4}, while we acknowledge that these calculations cannot completely rule out possible long-time-scale kinetic processes.
The electrostatic potential distribution on the (110) plane [Fig.~\ref{fig:md}(b)] further reveals that hydrogen atoms are strongly localized at the bottom of a deep potential well (-47 eV), suppressing proton diffusion and stabilizing the H sublattice~\cite{he2023phonon,he2024metal,tian2024few}.

We have evaluated the pressure-dependent enthalpies of the proposed \ce{PtH4}, \ce{PdH4}, and \ce{AuH4} phases (see Fig. S11). At ambient pressure all three compounds are thermodynamically unstable, lying approximately 191, 308, 432 meV/atom above the convex hull for \ce{PtH4}, \ce{PdH4}, and \ce{AuH4}, respectively. These distances from the convex hull indicate potential challenges for experimental synthesis under equilibrium or standard ambient conditions. 
We note, however, that our calculations may overestimate the actual distances above the convex hull due to the neglect of ionic quantum and anharmonic contributions, which can further stabilize superconducting hydrides~\cite{DANGIC2025100043,fang2025superconductivity,Errea2020-LaH10Nature}.
Moreover, anharmonic effects may also modify the phonon frequencies and eigenvectors, particularly for the hydrogen-dominated modes. In systems with pronounced phonon softening, such as \ce{H3S}~\cite{li2014metallization,duan2014pressure} and \ce{LaH10}~\cite{drozdov2019superconductivity,somayazulu2019evidence}, anharmonicity can lead to a reduction in the predicted \Tc\ by up to approximately 20\%.  We therefore emphasize that the calculated ($\lambda$) and \Tc\ values should be regarded as estimates within the harmonic approximation, and quantitative corrections arising from anharmonic phonon renormalization cannot be excluded. A reliable assessment of these effects would require fully anharmonic approaches, such as the stochastic self-consistent harmonic approximation (SSCHA) or related finite-temperature anharmonic lattice-dynamics methods, which are beyond the scope of the present work.

Furthermore, phases that are thermodynamically unfavorable under ambient conditions can possibly be realized under nonequilibrium or kinetically constrained conditions. For example, although Pt, Pd, and Au normally adopt the closely packed \textit{fcc} structure, high pressure and temperature can stabilize the \textit{bcc} phase~\cite{Cu-Ag-Pt-fcc-bcc-PRB2021,Pd-fcc-bcc-JAP2024}. Starting from a \textit{bcc} precursor may therefore facilitate the synthesis of the proposed $Im\bar{3}m$ MH$_4$. After formation, the pressure and temperature could be gradually reduced to kinetically stabilize the superconducting MH$_4$ at ambient pressure~\cite{gao2025maximum}. Alloying may further lower the required synthesis conditions. For instance, a \textit{bcc} TiFe-based hydride has been synthesized at only 5 GPa and 873 K by hydrogenating a TiFe alloy~\cite{bcc-TiFe-hydride2013IJHE}. Moreover, nonequilibrium approaches such as electron-beam irradiation can induce the $fcc$-$bcc$ transition in Au thin films at ambient pressure~\cite{Au-film-bcc-fcc-actmater2023}, offering an alternative route to synthesize MH$_4$ at ambient or near-ambient pressure.

We would like to clarify that, although the predicted
\ce{MH4} superconducting phases are thermodynamically metastable at ambient pressure, the primary purpose of this work is not
to claim equilibrium-stable ambient-pressure superconductors. Rather, the central contribution of our study is the development
and application of a chemical-template-driven strategy based on the interstitial charge characteristics of $bcc$ elemental metals to
search for novel hydrogen-rich superconductors. Using this approach, we identified several dynamically and thermally stable
hydrides exhibiting superconducting transition temperatures above the boiling point of liquid nitrogen. More importantly, this
design strategy is general and can be extended to other lattice types, potentially enabling the discovery of additional unconventional superconducting hydrides.
While the experimental realization of such phases remains challenging, they may become accessible under suitable non-equilibrium conditions such as high pressure, chemical precompression, irradiation-induced defects, or kinetic trapping, as well documented in the literature on hydrides~\cite{kataoka2022origin,li2024stabilization,liu2025high,shi2026ambient} and other metastable compounds~\cite{hinuma2016discovery,zhao2022fabrication,yu2019high,xu2024free}.

Beyond MH$_4$ (M = Pt, Pd, and Au), where eight hydrogen atoms occupy the positions of ISQs of the M lattice, we systematically explored the introduction of H$_n$ ($n = 1$-$7$) at these sites (see Figs. S12-S14). However, among these configurations, only PtH$_4$, PdH$_4$ and AuH$_4$ in the family of MH$_4$ were found dynamically stable.
Because Pt, Pd and Au all typically crystallize in the \textit{fcc} structure (space group $Fm\bar{3}m$) under ambient conditions, we also constructed hydrides based on the \textit{fcc} lattice. Phonon calculations were performed for the resulting structures, and only the dynamically stable ones were evaluated for EPC properties. Whereas, none of them exhibited a high \Tc\ (Figs. S14-S17). Of the hydrogenated \textit{fcc} Pt structures (i.e., PtH$_n$ in which $n$ runs from 1 to 8), only the $Pn\bar{3}m$ Pt$_4$H$_2$ is dynamically stable and exhibits a small \Tc~of 0.16 K. The incorporation of hydrogen atoms into \textit{fcc} Pd yields 10 dynamically stable structures. Among them, $Fm\bar{3}m$ Pd$_4$H$_8$ has a \Tc~of 6.8 K, and the \Tc~values of the rest hydrides are all below 3 K. In contrast, no dynamically stable structures are found when H atoms are injected into the \textit{fcc} lattice of Au. 

\begin{figure*}[htp]
	\centering
 \includegraphics[width=0.85\linewidth,angle=0]{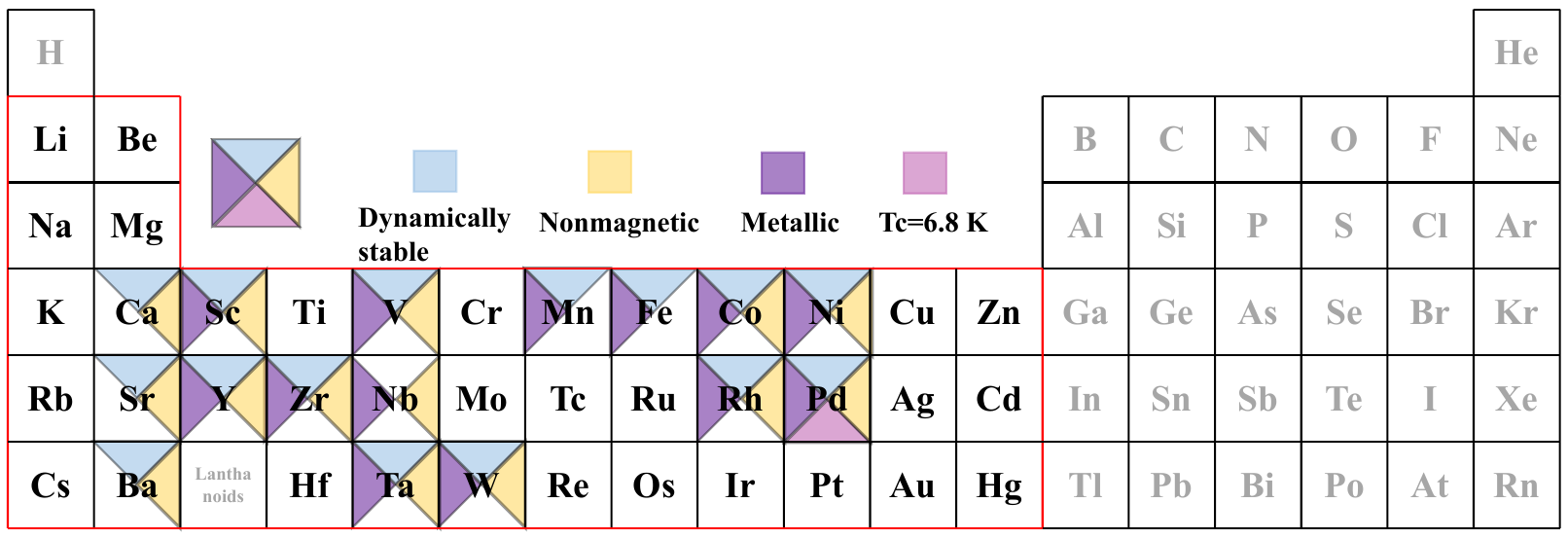}
	    \caption{\label{Fig:fcc-phonon} Results of high-throughput calculations of phonon
 dispersion curves, electronic properties, magnetic properties, and EPC calculations for M$_4$H$_8$}
\end{figure*}

Furthermore, we extended the $fcc$ \ce{M4H8} framework to other metallic elements beyond Pd, Au, and Pt (see Fig.5). In total, we identified another 16 dynamically stable compounds. One compound (\ce{FeH2}) is magnetic, and is therefore excluded from further consideration. Among the remaining compounds, six hydrides Sc/V/Y/Nb/Ta/ZrH$_2$ have been experimentally synthesized~\cite{venturini1977low,khodyrev1980electron,xiao2009energetics,nowak1984nuclear,muller1986investigation,maeland1961novel,pebler1962crystal,burnasheva1984formation,wang1994energetics,markin1983erbium,uhrich1966measurement,beck1962zirconium,pitt2011situ,sidhu1962neutron}, three materials including Ca/Sr/BaH$_2$ are nonmetallic, and five compounds are dynamically stable and metallic. However, except for \ce{Pd4H8}, none of these compounds exhibits superconductivity.

\section{CONCLUSIONS}
In summary, we report a material family of high-\Tc~ hydrides, MH$_4$ (M = Pt, Pd, Au), stabilized at ambient pressure in the PtHg$_4$-type  structure—a previously unexplored structural motif for hydride superconductivity, in which hydrogen atoms occupy the one-quarter body-diagonal sites of the metal lattice. Calculations of the electronic band structures, phonon spectra, and electron-phonon coupling predict \Tc~of 84~K, 133~K, and 89~K for PtH$_4$, PdH$_4$, and AuH$_4$, respectively. The large EPC parameter $\lambda_H$ and the pronounced phonon softening originate from hydrogen vibrations and strong Fermi surface nesting, which collectively enhance the superconductivity. Beyond high-throughput screening, our work provides mechanistic insight linking superconductivity to specific phonon softening at the H point, offering clear guidance for designing future ambient-pressure hydride superconductors by targeting similar mechanisms. These findings open a viable pathway toward ambient-pressure high-\Tc\ superconductivity and suggest promising materials as touchstones for future experimental validation.

\section{Acknowledgments}

The authors thank A. Sanna for insightful discussions on the concepts of SCDFT and Migdal-Eliashberg theory.
The authors acknowledge funding from the NSFC under grants Nos. 12674011, 12474012,  12574064, 12174160 and 12074154. W.C., and M.A.L.M. acknowledge the funding from the Sino-German Mobility Programme under Grant No. M-0362. Y.L. and J.S. acknowledge the funding from 333 High-level Talents Project of Jiangsu Province. W.C., Y.L., and H.L. acknowledge the funding from the Open Project of State Key Laboratory of High Pressure and Superhard Materials (Jilin University) (No. 202513).
X.L. acknowledges the funding from Postgraduate Research \& Practice Innovation Program of Jiangsu Province NO. 2025XKT0776. All the calculations were performed at the High Performance Computing Center of the School of Physics and Electronic Engineering of Jiangsu Normal University.
Y.-W.F. acknowledges Extraordinary Grant of CSIC (Grant No. 2025ICT122) and the financial support received from the IKUR Strategy under the collaboration agreement between Ikerbasque Foundation and Centro de Física de Materiales (CFM-MPC) on behalf of the Department of Science, Universities and Innovation of the Basque Government (HPCAI21: AI-CrysPred).
Y.-W.F. acknowledges enlightening discussions with his colleagues including D. Dangi\'c and I. Errea.
M.A.L.M. was supported by a grant from the Simons Foundation (SFI-MPS-NFS-00006741-12, P.T.) in the Simons Collaboration on New Frontiers in Superconductivity and by the Kavli and the Klaus Tschira foundations as a part of the SuperC collaboration.

\section{Author Contributions}

 W.C., Y.L.,  Y.-W.F. and M.A.L.M. conceived the project. W.C., Y.L. and Y.-W.F. supervised the research. X.L., W.X, Y.-W.F., W.C. and M.A.L.M. performed the simulations and analyzed the data. X.L., W.X, Y.-W.F., Y.L., M.A.L.M. and W.C. co-wrote the manuscript; all authors discussed the results and commented on the manuscript.

\section{Data availability}

The data supporting this study's findings are available within the article.

\section{Declaration of interests}
The authors declare no competing interests.

% \bibliography{MH4}

\end{document}